\documentclass[11pt]{article}
\usepackage{mathptmx}
\usepackage{tabularx}
\usepackage{booktabs}
\usepackage{multirow}
\usepackage{amsmath}
\usepackage{amsfonts}
\usepackage{fancyhdr}
\usepackage{graphicx}
\usepackage{hyperref}
\usepackage{cite}
\usepackage{float}
\usepackage{color}
\usepackage[margin=1in]{geometry}
\usepackage{blindtext}
\usepackage{appendix}
\usepackage{graphicx}
\usepackage{subcaption}
\usepackage[utf8]{inputenc}
\usepackage[export]{adjustbox}
\usepackage{wrapfig}
\date{}
\begin{document}
\title{\textbf{Dynamics of General Barotropic Stellar Fluid in the Framework of $R+2\alpha T$ Gravity}}
\maketitle
\begin{center}
\author{Karim Mosani$^{1}$, Gauranga C. Samanta$^{2}$ \\
Department of Mathematics, BITS Pilani K K Birla Goa Campus, Sancoale, Goa 403726, India.\\
$^1$kmosani2014@gmail.com \\
$^2$gauranga81@gmail.com
}

\end{center}

\begin{abstract}
Gravitational collapse of a spherically symmetric homogeneous perfect barotropic fluid with linear as well as polytropic type Equation of State (EoS) has been investigated in the framework of a linear model of  $f(R,T)$ gravity. This modified gravity has the potential to explain the observed cosmic acceleration. The calculations have been done taking the transformed time coordinate $t \to \sqrt{\frac{\rho_0}{3}} t$, where $\rho_0$ is the initial density of the fluid. For linear EoS $p=\omega\rho$,
%where $\omega$ is the EoS parameter, 
the condition for being a true singularity, along with sufficient condition for the  formation of apparent horizon covering the singularity has been derived. For a polytrope having the EoS $p=K\rho^{1+\frac{1}{n}}$, the scale factor ($A$) as a function of fluid density ($\rho$) has been obtained which is then used to study the dynamics of the fluid. Role of the polytropic index ($n$) and the constant of proportionality ($K$) in the dynamics of the fluid is also studied. A new type of exotic matter field having varied dependence of scale factor on the density, and having the potential to give rise to bouncing cosmology, provided it is the dominating fluid in the universe, is obtained in this domain and is investigated. Energy conditions are discussed. 
\end{abstract}

{\bf \it{Key words}:} Cosmic Acceleration, $f(R,T)$ Gravity, Gravitational Collapse, Singularity.

\section{Introduction}
Early $20^{th}$ century has seen major breakthroughs in the field of cosmology.  To visualize the space time as a Riemannian manifold which can be stretched or twisted, and the formulation of gravity in terms of curvatures in this fabric of cosmos has changed the way we look at the structures in the universe. The idea of gravity as a geometric property of spacetime was proposed by Einstein \cite{Weinberg} in 1915 and is put wonderfully in the equation,
$$
G_{\mu \nu}=\frac{8\pi G}{c^4}T_{\mu \nu}.
$$
where $G_{\mu \nu}$ and $T_{\mu \nu}$ are the components of the Einstein tensor and the stress-energy tensor respectively. 

Recent observations of the High-Z Supernova Search Team \cite{Riess} and the type Ia supernova by the Supernova Cosmology Project \cite{Perlmutter}  suggests that the universe  is undergoing cosmic acceleration. These observations are not at par with the the outcome of general relativity (GR)  which predicts a positive deceleration parameter \cite{Weinberg, Liddle}, or in other words, the decelerated expansion of the universe, provided the universe is completely filled with ordinary matter, the one which satisfies all the four energy conditions \cite{Hawking}.

Existence of dark energy models, which are exotic forms of matter field, however not satisfying the energy conditions  and having the repulsive property, have been considered to explain the cosmic acceleration all the while maintaining the uprightness of GR. These models include the cosmological constant \cite{Einstein}, quintessence \cite{Chiba2}, k-essence \cite{Picon2}, phantom energy \cite{Singh} to name a few. A detailed discussion about the dark energy models   could be found in \cite{Amendola}.

Alternatively, modifications in the Lagrangian of Einstein-Hilbert (EH) action has been proposed in which, instead of the Ricci scalar ($R$), some more general function like the $f(R)$ \cite{Starobinsky}, $f(R,T)$ \cite{Harko1} come into the picture. Apart from this, the Gauss-Bonnet (GB) gravity \cite{Nojiri} and the scalar tensor theories like the Brans-Dicke (BD) theory \cite{Brans}  have also been used to match with the current observations.

Coming to the aspects of the behavior of a massive star, once the nuclear fuel supply, which keeps the star in equilibrium by balancing the inward gravitational pull, is over, the gravitational effect dominates and the star is pulled inwards indefinitely. This could give rise to the occurrence of singularities which is either  hidden behind the apparent horizon giving rise to a black hole, or naked, with the ability to communicate with the outside world. Detailed information about the nature of singularity could be found in \cite{Joshi}.   In this domain of strong gravity pull, the model of gravity could play a significant role in the outcome of the collapse.

The motivation of the present work is to understand the stellar dynamics under the influence of a linear model of $f(R,T)$ gravity in which the Lagrangian is $R+2\alpha T$.  This model whose more general version was investigated by Poplawski \cite{Poplawski} as a cosmological model having varying cosmological term $\Lambda$ being a function of the trace of stress energy tensor ($T$).  In our particular case, we have $\Lambda (T)=-\alpha T$. This model has equivalence with a GR model plus a cosmological term which is dynamic in nature, i.e. $\Lambda \propto A^{-2}$ as discussed in \cite{Chen} or $\Lambda \propto H^2$ as discussed in \cite{Lima}. Here, $H$ is nothing but the Hubble constant having relation with $A$ as $H=\frac{\dot A}{A}$. Having a $\Lambda$ which is decreasing with time has an upper hand over the constant cosmological term that it it has the potential to explain the large entropy of the universe which is observed. Also variable cosmological term is favourable by other observed cosmological data like the WMAP \cite{Spergel} and the SNIa gold sample data  \cite{Riess2} , as discussed in \cite{Poplawski}.
%As it is put wonderfully by Prof. Joshi, "This is the arena where we in fact come face to face with the regime of extreme and ultra strong gravity fields, and the answers must be sought within the framework of a gravitation theory such as the Einsteins theory of gravity".
 
 Organization of the paper is as follows. Modified field equations and the corresponding general Equation of Continuity (EoC) in $f(R,T)$ gravity is discussed in Section ($2$). Equations governing the stellar perfect fluid in presence of EH action having Lagrangian $R+2 \alpha T$ is derived in Section ($3$). Equations derived in this section are applied to investigate the dynamics of the stellar barotropic perfect fluid having linear EoS $p=\omega \rho$ in Section ($4$). Corresponding nature of singularities formed are also discussed in this section. Dynamics of polytrope in the given model of gravity is discussed in Section ($5$). Results obtained are summarized in Section ($6$).

\section{Mathematical Formulation for $f(R,T)$ Gravity}
In $f(R,T)$ gravity, the Lagrangian in the Einstein-Hilbert (EH) action is  an arbitrary function of Ricci scalar ($R$) and trace ($T$) of the stress-energy tensor of matter ($T_{\mu \nu}$). The full action is then given by sum of the EH action and the term describing the matter field as
\begin{equation}
    S= \frac{1}{2}\int f(R,T) \sqrt{-g} d^4x +\int \mathcal{L}_m \sqrt{-g}d^4x,
\end{equation}
where $\mathcal{L}_m$ is the matter Lagrangian density. Here we use the scale $c=8\pi G = 1$. The stress-energy tensor is defined in terms of $\mathcal{L}_m$ as
\begin{equation} \label{setl}
    T_{\mu \nu} = -\frac{2}{\sqrt{-g}}\frac{\delta (\sqrt{-g}\mathcal{L}_m)}{\delta g_{\mu \nu}}.
\end{equation}
Variation of the full action with respect to $g^{\mu \nu}$ leads to the following gravitational field equation upon simplification:
\begin{equation}
    \begin{split}
    & f_R (R,T)\Big [ R_{\mu \nu}-\frac{1}{3}R g_{\mu \nu} \Big ]  +\frac{1}{6} f(R,T) g_{\mu \nu}=[1-f_T(R,T)] \Big [ T_{\mu \nu} - \frac{1}{3}T g_{\mu \nu} \Big]-f_T(R,T) [\Theta _{\mu \nu}-\frac{1}{3}\Theta g_{\mu \nu}] \\
    & + \nabla _\mu \nabla _\nu f_R(R,T),
\end{split}
\end{equation}
where $\Theta _{\mu \nu}= g^{\alpha \beta}\frac{\delta T_{\alpha \beta}}{\delta g_{\mu \nu}}$ and $\Theta=\Theta _{\mu}^{\mu}$. The equation of continuity followed in the framework of $f(R,T)$ gravity has a general form given by
\begin{equation} \label{eoc}
    \nabla ^\mu T_{\mu \nu}=\frac{f_T(R,T)}{1- f_T(R,T)} [ [T_{\mu \nu} +\Theta_{\mu \nu}]\nabla^\mu \ln{f_T(R,T)}+\nabla^\mu \Theta_{\mu \nu} ].
\end{equation}
This equation is obtained by taking the co-variant divergence of the field equation and using the identity
\begin{equation}
    \nabla^\mu \Big [ f_R(R,T)R_{\mu \nu}- \frac{1}{2} f(R,T) g_{\mu \nu} +[g_{\mu \nu} \Box -\nabla_{\mu} \nabla_\nu] f_R(R,T) \Big ]=0.
\end{equation}
$\Theta _{\mu \nu}$ is now represented in terms of $\mathcal{L}_m$ by differentiating  Eq.(\ref{setl}) and using the fact that $\mathcal{L}_m$ depends only on $g_{\mu \nu}$ and not on its derivative thereby giving us the relation 
\begin{equation} \label{thet}
    \Theta _{\mu \nu}= -2 T_{\mu \nu} +g_{\mu \nu} \mathcal{L}_m -2g^{\alpha \beta} \frac{\partial ^2 \mathcal{L}_m}{\partial g^{\mu \nu} \partial g^{\alpha \beta}}.
\end{equation}
We now put into use the results obtained in this section in a particular model of $f(R,T)$ gravity discussed in the next section. 
\section{Collapse dynamics}
We consider a simple model given by $f(R,T)=R+ 2 f(T)$, where $f(T)=\alpha T$  ($\alpha$ is a constant), and analyze the collapse scenerio of a perfect fluid whose stress energy tensor components $T^\mu_\nu$ are given by
\begin{equation*}
    T^\mu_\nu= diag(\rho,-p,-p,-p).
\end{equation*}
Usin Eq.(\ref{thet}), the relation for the above matter field is obtained as 
\begin{equation*}
\Theta _{\mu \nu}=-2 T_{\mu \nu} -p g_{\mu \nu}.    
\end{equation*}
The field equation then becomes
\begin{equation}
G^\mu_\nu=T^\mu_\nu+2f_T(T) T^\mu_\nu+[2 p f'(T) + f(T)]g^\mu_\nu.
\end{equation}
 We  assume the metric components of the spherically symmetric metric, governing the interior of the star, to be separable in terms of functions of $r$ and $t$, also known as Lemaitre-Tolman-Bondi (LTB) metric, as follows:
\begin{equation}
    ds^2=dt^2-A^2(t)h(r)dr^2-A^2(t)r^2d\Omega^2,
\end{equation}
where
\begin{equation*} \label{metric}
    d\Omega^2=d\theta^2+\sin^2{\theta}d\phi^2.
\end{equation*}
The following field equations for this type of metric is realized:
\begin{align}
& \frac{3\dot A^2}{A^2}+\frac{h'}{A^2h^2r}-\frac{1}{A^2hr^2}+\frac{1}{A^2r^2}=[1+3\alpha]\rho- \alpha p, \label{0,0} \\
& \frac{\dot A^2}{A^2}+\frac{2\ddot A}{A} -\frac{1}{A^2hr^2}+\frac{1}{A^2r^2}=\alpha \rho -[1+3\alpha] p, \label{1,1} \\ 
& \frac{\dot A^2}{A^2}+\frac{2\ddot A}{A}+\frac{h'}{2A^2h^2r}=  \alpha \rho - [1+3\alpha] p.  \label{2,2}
\end{align}
    From Eq.(\ref{1,1}) and Eq.(\ref{2,2}), we get
    \begin{equation}
        \frac{h'}{2h^2r}+\frac{1}{hr^2}-\frac{1}{r^2}=0,
    \end{equation}
    which is satisfied by
    \begin{equation} \label{hr}
        h(r)=\frac{1}{1+C_1 r^2}.
    \end{equation}
    where $C_1$ is a constant. This is just like the one which is obtained in the framework of GR.
    Hence, we conclude that assuming the metric of the form [\ref{metric}], we get the Friedmann-Lemaitre-Robertson-Walker (FLRW) metric 
    \begin{equation} \label{flrw}
         ds^2=dt^2-\frac{A^2(t)}{1+Cr^2}dr^2-A^2(t)r^2d\Omega^2.
    \end{equation}
    Substituting the value of $h(r)$ from Eq.(\ref{hr}) in the first field equation Eq.(\ref{0,0}) gives
    \begin{equation} \label{0,0-2}
        \frac{\dot A^2}{A^2}- \frac{C_1}{A^2}= \Big [ \frac{1}{3} +\alpha \Big ] \rho - \alpha p. 
    \end{equation}
We can assume without loss of generality that  $A(0)=1$ and $A'(0)=0$, which means that the scale factor has been normalised to unity initially, to get the value of integration constant $C_1$ as
\begin{equation} \label{c}
    C_1=\frac{\alpha p_0-[3\alpha+1]\rho_0}{3} ,
\end{equation}
where $\rho_0$ is the initial density of the collapsing body. Using this in the Eq.(\ref{0,0-2}) gives 
\begin{equation} \label{0,0-3}
    \dot A^2-\Big [\frac{1}{3}+ \alpha \Big ][A^2\rho -\rho_0]+\frac{\alpha}{3}[A^2p-p_0]=0.
\end{equation}
From the field equations Eq.(\ref{0,0}) and Eq.(\ref{1,1}) along with using Eq.(\ref{c}), the following equation, sometimes known as the acceleration equation, is obtained:
\begin{equation} \label{acceleration equation}
    \frac{\ddot A}{A}=-\frac{1}{6}[\rho+3p]-\frac{1}{6A^2}[\alpha p_0-[3\alpha+1]\rho_0].
\end{equation}
If the RHS of the above equation is positive, it implies that the dynamics is accelerated. Similarly negative RHS implies decelerated change. Eq.(\ref{acceleration equation}) in absence of the second term in RHS is nothing but the acceleration equation corresponding to the entire universe governed by FLRW metric with negligible curvature. Presence of this particular term signifies the satisfaction of the initial conditions $A(0)=1$ and $\dot A(0)=0$ making the equation viable for our stellar models.

To find how the density of the body ($\rho$) varies with the scale factor ($A(t)$) we use the equation of continuity in this particular case, which is obtained by using Eq.(\ref{eoc}), as:
\begin{equation} 
    \dot \rho +\Big [\frac{2\alpha}{1+2\alpha} \Big ]\dot p +\frac{3\dot A}{A}[\rho +p]=0.
\end{equation}

\section{Barotropic Linear Equation of State}

\begin{figure}\begin{center} 
\includegraphics[scale=0.5]{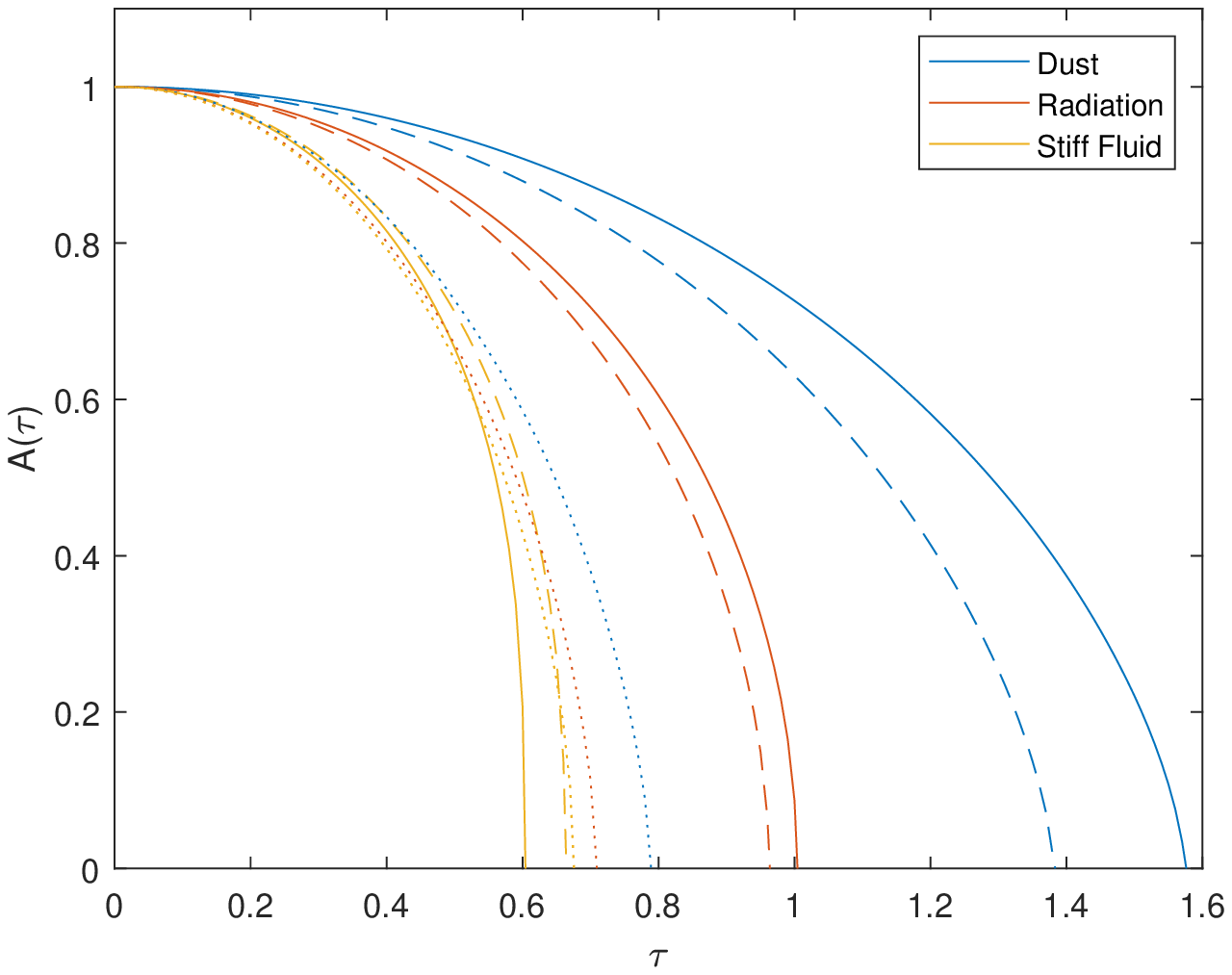} \includegraphics[scale=0.5]{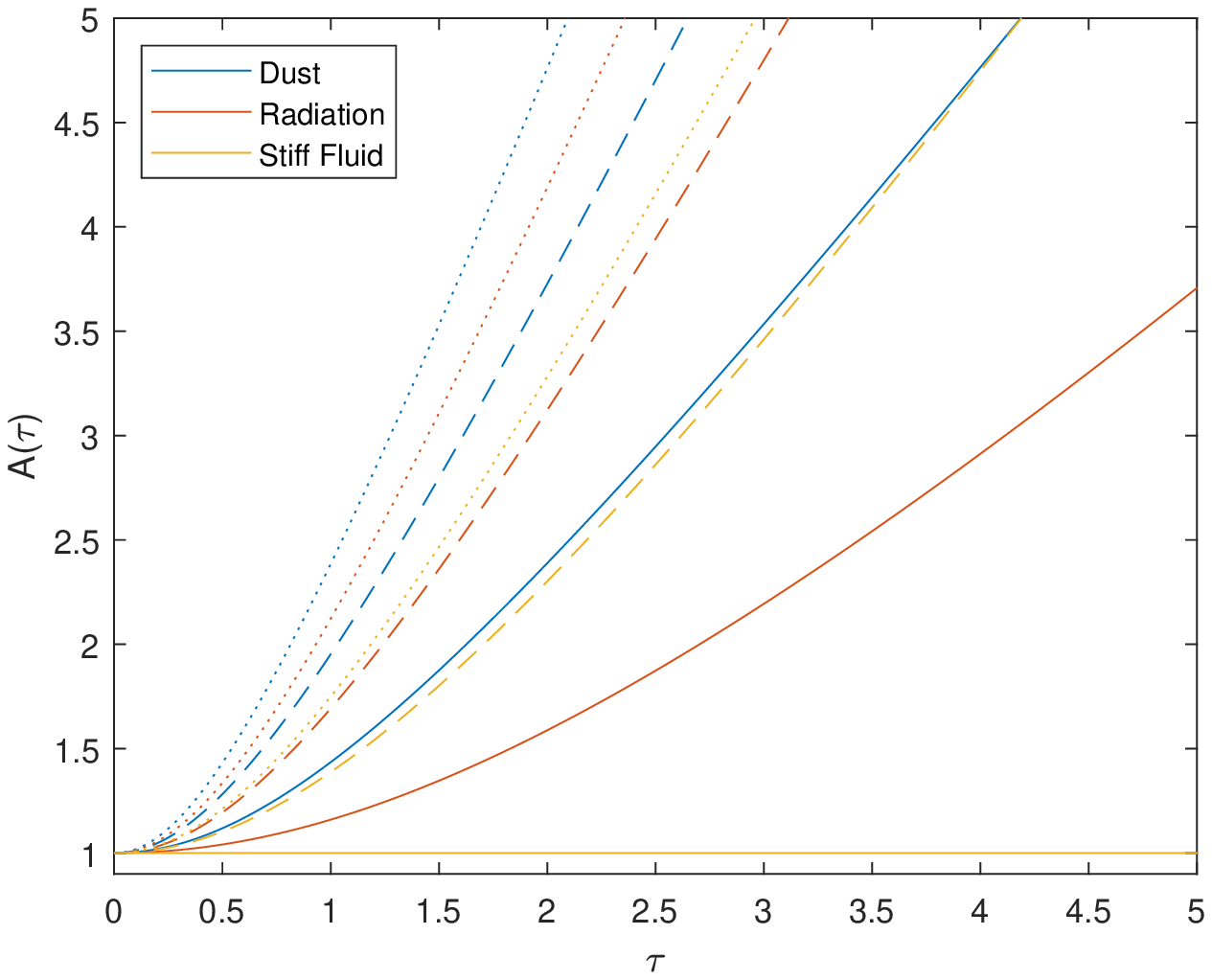}  
\caption{Dynamics of scale factor $A(\tau)$ for various barotropic fluid with linear EoS in presence of modified gravity having Lagrangian $R+2\alpha T$, where the solid, dashed and dotted lines represent models with $\alpha=0$, $0.1$ and $1$ respectively in the left panel and $\alpha=-1$, $-2$ and $-3$ respectively in the right panel.  } \label{sf1}
\end{center}
\end{figure}

Assuming linear Equation of State (EoS), $P=\omega \rho$, where $\omega$ is the EoS parameter, and using the above equation we get a particular EoC as
\begin{equation}
    \dot \rho +\frac{3[1+\omega][1+2\alpha]}{[1+2\alpha[1+\omega]]} \frac{\dot A}{A} \rho =0,
\end{equation}
giving the relation between $\rho$ and $A$ as
\begin{equation} \label{eoc1}
    \rho= \rho_0 A \string^ \Big [-\frac{3[1+\omega][1 +2\alpha]}{1 + 2\alpha[1 + \omega]}  \Big].
\end{equation}
One could observe that for a pressureless fluid, also known as dust ($\omega=0$), density dependence on scale factor is free of $\alpha$ and is simply $\rho=\rho_0 A^{-3}$. Particular case for GR is obtained corresponding to $\alpha=0$ as $\rho=\rho_0 A^{-3[1+\omega]}$, as expected. Using Eq.(\ref{0,0-3}) and Eq.(\ref{eoc1}), and introducing dimensionless time ($\tau$) obtained from the transformation $t \to t \sqrt{\frac{\rho_0}{3}}  (=\tau)$, we get the following differential equation governing the dynamics of the collapsing star:
\begin{equation} \label{collapse de}
  \frac{dA}{d\tau}=\pm \sqrt{[\alpha[\omega-3]-1]\Big [  1- A \string^ \Big [ -1-\frac{3\omega}{1+2\alpha[1+\omega]} \Big ]}.
 \end{equation}
 Thus, two differential equations, differing in polarity,  are obtained out of  which only one governs the dynamics of the fluid. For a particular type of fluid (particular value of $\omega$) and a particular linear model (fixed value of $\alpha$), the differential equation which has a real solution is considered and the one having complex solution with non-zero imaginary component  is discarded. For a given linear model $R+ 2\alpha T$ of $f(R,T)$ gravity, the fluid having linear EoS parameter $\omega=\frac{3\alpha+1}{\alpha}$ or $\omega=\frac{2\alpha+1}{2\alpha+3}$ does not have any dynamics involved and is static throughout.
When $\alpha \to 0$, the Eq.(\ref{collapse de}) reduces to 
\begin{equation}
\frac{dA}{d\tau}= \pm \sqrt{A^{-[1+3\omega]}-1}.
\end{equation}
This is nothing but the one corresponding to the dynamics of gravitational collapse in the frame work of GR which has been investigated in \cite{Astashenok}. Dynamics of scale factor for dust ($\omega=0$), radiation (which includes particles moving at relativistic speed having $\omega=\frac{1}{3}$)  and stiff fluid (which attains the largest possible value of EoS parameter ($\omega=1$) without violating causility, i.e. the sound speed is same as the speed of light in this fluid \cite{stiff}), and the influence of $\alpha$ has been plotted in Figure \ref{sf1}. It is also observed that pressureless fluid collapses in the gravity model for which $\alpha>-\frac{1}{3}$ and expands in the model for which $\alpha<-\frac{1}{3}$. Similarly, relativistic particles collapses (expands) in the model for which $\alpha>-\frac{3}{4}$ ($\alpha<-\frac{3}{4}$) and for stiff fluid, the domain for collapse (expansion) is $\alpha>-1$ ($\alpha<-1$). 
 
\subsection{Nature of singularity}
Singularities occurring in space-time could either be a true one, which is physically significant, or it could be arising due to mathematical pathologies. In the latter case,  diffeomorphism invariant quantities like the space-time curvature remains finite along the timelike or null incomplete geodesic ending at the singularity. This pseudo singularities could be eliminated by changing the coordinate system. The Ricci scalar ($R$) and the Kretschmann scalar ($K$), which is the square of Riemann tensor given by $R_{\mu \nu \gamma \delta} R^{\mu \nu \gamma \delta}$,  corresponding to the metric represented by Eq.(\ref{flrw}) are given by 
\begin{equation}
    R=\frac{6\ddot A}{A} +\frac{6 \dot A^2}{A^2}- \frac{2[[3\alpha +1]\rho_0 -\alpha p_0]}{3A^2}
\end{equation}
and
\begin{equation}
    K=\frac{12 \Big [ A^2\ddot A^2 +\Big [ \frac{[3\alpha +1]\rho_0 -\alpha p_0}{3}+\dot A^2 \Big ]^2  \Big ] }{A^4}.
   \end{equation}
Substituting the value of $\dot A$ and $\ddot A$ in terms of $A$ using Eq.(\ref{collapse de}) in the above equations gives us
\begin{equation}
  %  R=-6\sqrt{\frac{\rho_0}{3}}\zeta \eta A^{\eta-2}+ \frac{2\rho_0 \zeta [1-A^{\eta}]}{A^2}+\frac{2 \rho_0 \zeta}{3A^2}
  R=-2\zeta \rho_0 \Bigg [ \sqrt{\frac{3}{\rho_0}}\eta+1 \Bigg ] A^{\eta-2}+\frac{8\rho_0\zeta}{3A^2}
\end{equation}
and
\begin{equation}
   % K=\zeta^2 \rho_0 \Big [ 4\eta^2 A^{2\eta-4}+\frac{4 \rho_0}{3 A^4}-\frac{8\rho_0 [1-A^{\eta}]}{3A^4} +\frac{4\rho_0[1-A^{\eta}]^2}{3A^4} \Big ],
   K=\Big [ 4\zeta^2\eta^2\rho_0+ \frac{4\rho_0}{3} \Big ] A^{2[\eta-2]},
\end{equation}
where 
\begin{equation} \label{zetaeta}
    \zeta= \alpha[ \omega-3]-1, \quad \eta=-\frac{3\omega}{1+2\alpha[1+\omega]}-1.
\end{equation}
%$\zeta= \alpha[ \omega-3]-1$ and $\eta=-\frac{3\omega}{1+2\alpha[1+\omega]}-1$.
For non-zero $\zeta$, $R$ blows up whenever $A\to 0$. It is noted that $\zeta=0$ corresponds to a fluid having linear EoS parameter $\omega=\frac{3\alpha+1}{\alpha}$ which is static for a given $\alpha$.  Also for $\eta<2$, $K$ blows up as $A\to 0$. There are certain examples of singularities which are merely mathematical pathologies, for e.g. when $(\alpha,\omega)=(-\frac{1}{3},-1)$, ($\eta=2$), in which case $K$ remains constant and does not blow up as $A\to 0$. Hence, for $\zeta \neq 0$ and $\eta<2$, the singularities formed are true and physical because of blow up of both the Ricci scalar and the Kretschmann scalar.
%In case of collapse of a pressureless fluid, $\eta=-1$ and it is noted that both $R$ and $K$ blows up to infinity as $A \to 0$ indicating that the singularity thus formed is indeed a physical one and not merely a removable one.

Visibility of singularity depends on the time of formation of the Apparent Horizon (AH) \cite{AHref} , which is a hypersurface covering the central shell focussing singularity cutting its contact with the outside observer by trapping all the outgoing non-space like geodesics. It is basically a dynamic analog of an event horizon (EH). For a static star, EH and AH coincides. If the singularity forms prior to the development of AH, then we could obtain a naked singularity, thereby violating the cosmic censorship hypothesis which states that singularity is not visible from future null infinity. The condition required to form the AH is 
\begin{equation} \label{ah}
    g^{\mu \nu} Y,_\mu Y,_\nu =0,
\end{equation}
where $Y$ is the proper radius of the 2-sphere. Eq.(\ref{ah}) for $Y=rA$ then becomes
\begin{equation}
    \dot A^2=\frac{1}{r^2}+ \frac{\rho_0}{3}[\alpha[\omega-3]-1] =\delta^2,
\end{equation}
where $\delta ^2$ has been introduced, which is a positive constant, since the LHS and RHS of the first equation are functions of $t$ and $r$ respectively, and are hence constant.
Using the above equation along with  Eq.(\ref{collapse de}), the scale factor during the time of formation of the AH is obtained as
\begin{equation}
  %  A(t_{AH})=\Big [ \frac{3\delta ^2 +\rho _0[1 + \alpha [3-\omega]}{\rho _0 [1+\alpha [3-\omega]]} \Big ] \string^ \Big [ -\frac{1+2\alpha [1+ \omega]}{1+2\alpha [1+ \omega] +3 \omega}  \Big]
A(t_{AH})=\Bigg [ 1-\frac{\delta^2}{\zeta} \Bigg]^{\frac{1}{\eta}}.
\end{equation}
%In the case of dust collapse in the framework of the cosmological model corresponding to Lagrangian $f(R,T)=R+ 2\alpha T$, singularity is formed only in the sub case which satisfies $\alpha>-\frac{1}{3}$. Substituting $\omega=0$ in the above equation yields
%\begin{equation} 
 %   A(t_{AH})= \frac{\rho_0 [1+3\alpha]}{\delta^2 +\rho_0 [1+3\alpha]}
%\end{equation}
A sufficient condition for formation of AH before the formation of singularity is to have $\delta^2 < \zeta$. This is because $A(t)\geq 0$ and $\dot A(t)<0$, making the scale factor a monotone decreasing function of $t$. Also, the restrictions $\zeta \neq 0$ and $\eta<2$ has to be maintained for the singularity to be physical. A special case is obtained when $\delta^2=\zeta$, giving $A(t_{AH})=0$ and implying that the AH and the singularity forms concurrently.
 %Since $A(t)\geq 0$ and only the collapsing solution ($\dot A(t)<0$) is taken into account, we could safely say that the singularity is hidden behind the apparent horizon and that the cosmic censorship hypothesis is satisfied. 

\section{Barotropic Non Linear EoS: Polytropes}
Polytropic star is the solution of what is known as Lane-Emden equation, having non linear EoS of the form
\begin{equation}
    p=K \rho^{1+\frac{1}{n}},
\end{equation}
where $K$ is a constant whose value depends on the chemical composition of the star and its entropy per nucleon, and $n$ is called the polytropic index \cite{Weinberg}. This form of EoS arises due to the supposition that convection is the most effective way of energy transfer inside a supermasive star \cite{Jun}. The EoC for the above EoS in $R+ 2\alpha T$ gravity is derived from Eq.(\ref{eoc}) as 
\begin{equation}
    \dot \rho +\frac{3[1+2\alpha]}{1+2\alpha[1+[1+\frac{1}{n}] K \rho^{\frac{1}{n}} ]}\rho[1+K \rho^{\frac{1}{n}}]\frac{\dot A}{A}=0.
\end{equation}
Solving this we get
\begin{equation} \label{eoc2}
    A(\rho)=\Bigg [ \frac{\rho}{\rho_0}\Bigg ]^{-\frac{1}{3}} \Bigg [\frac{1+K \rho ^{\frac{1}{n}}}{1+K \rho_0^{\frac{1}{n}} } \Bigg ]^{\frac{2\alpha -n}{-3[2\alpha +1]}}.
\end{equation}
It is noted that a linear EoS corresponds to polytropic index $n \to \infty$ and $K=\omega$. Letting $n \to \infty$ in the above equation and taking the limit gives
\begin{equation}
    A(\rho)=\Bigg [ \frac{\rho}{\rho_0}\Bigg ]^{ -\frac{1}{3}} \exp{ \Bigg [\frac{\omega}{3[1+ \omega]}\log \Bigg [ \frac{\rho}{\rho_0} \Bigg ] \Bigg ]}.   
\end{equation}
Simplifying this gives Eq.(\ref{eoc1}) as expected. Now substituting for scale factor from Eq.(\ref{eoc2}) in Eq.(\ref{0,0-3}) gives the following non linear first order differential equations for $\rho$:
\begin{equation} \label{polytropecollapsedynamics}
    \frac{d\rho}{d\tau}=\pm \frac{\sqrt{\text{F}_2(\rho)+\text{F}_c}}{\text{F}_1(\rho)},
\end{equation}
where
\begin{equation*}
\begin{split}
    & \text{F}_1 =\left |\frac{1}{3}\Bigg [\frac{1+K\rho^{\frac{1}{n}}}{1+K\rho_0^{\frac{1}{n}}}\Bigg ]^{-\frac{[2\alpha-n]}{3[2\alpha+1]}} \Bigg [ \frac{\rho}{\rho_0}\Bigg ]^{-\frac{1}{3}} \Big [ 1-\frac{[2\alpha-n]K \rho^{ [ -1+\frac{1}{n}  ]}}{n[2\alpha+1][1+K\rho^{\frac{1}{n}}]} \Big ]\right |,  \\
    &  \text{F}_2 =\frac{1}{3}\Big [\frac{\rho}{\rho_0}\Big ]^{-\frac{2}{3}} \Bigg [\frac{1+ K \rho^{\frac{1}{n}}}{1+K\rho_0^{\frac{1}{n}}} \Bigg ]^{\frac{2[2\alpha-n]}{3[2\alpha+1]}}[1+3\alpha[1-K\rho^{\frac{1}{n}}]]\rho, \\
    & \text{F}_c  =\frac{1}{3}[\alpha[K\rho^{\frac{1}{n}}-3]-1]\rho_0.
     \end{split}
 \end{equation*}

\begin{figure}
\begin{subfigure}{.5\textwidth}
  \centering
 \includegraphics[scale=0.57]{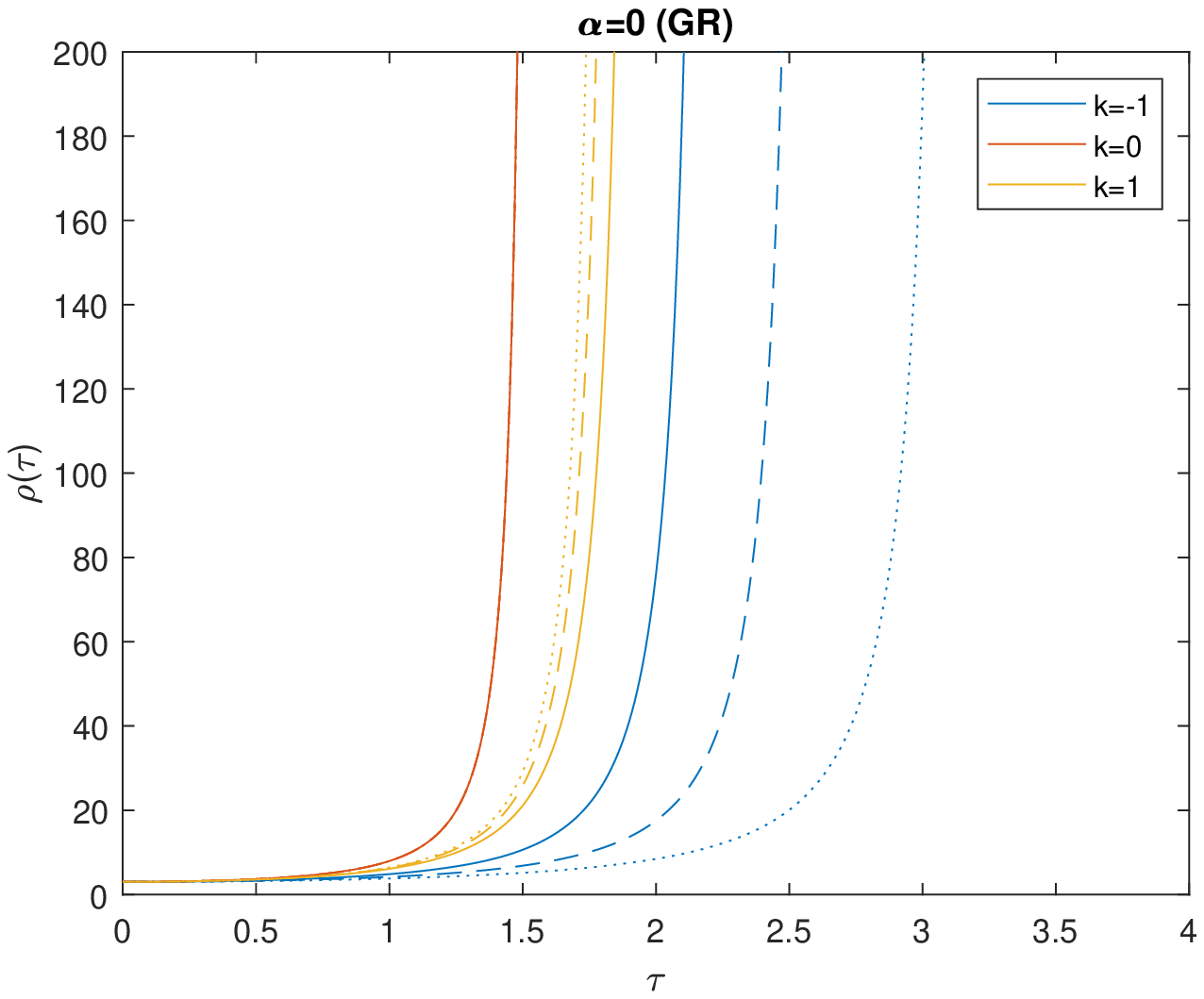}
  \caption{}
  \label{fig2a}
\end{subfigure}%
\begin{subfigure}{.5\textwidth}\label{fig2b}
  \centering
 \includegraphics[scale=0.57]{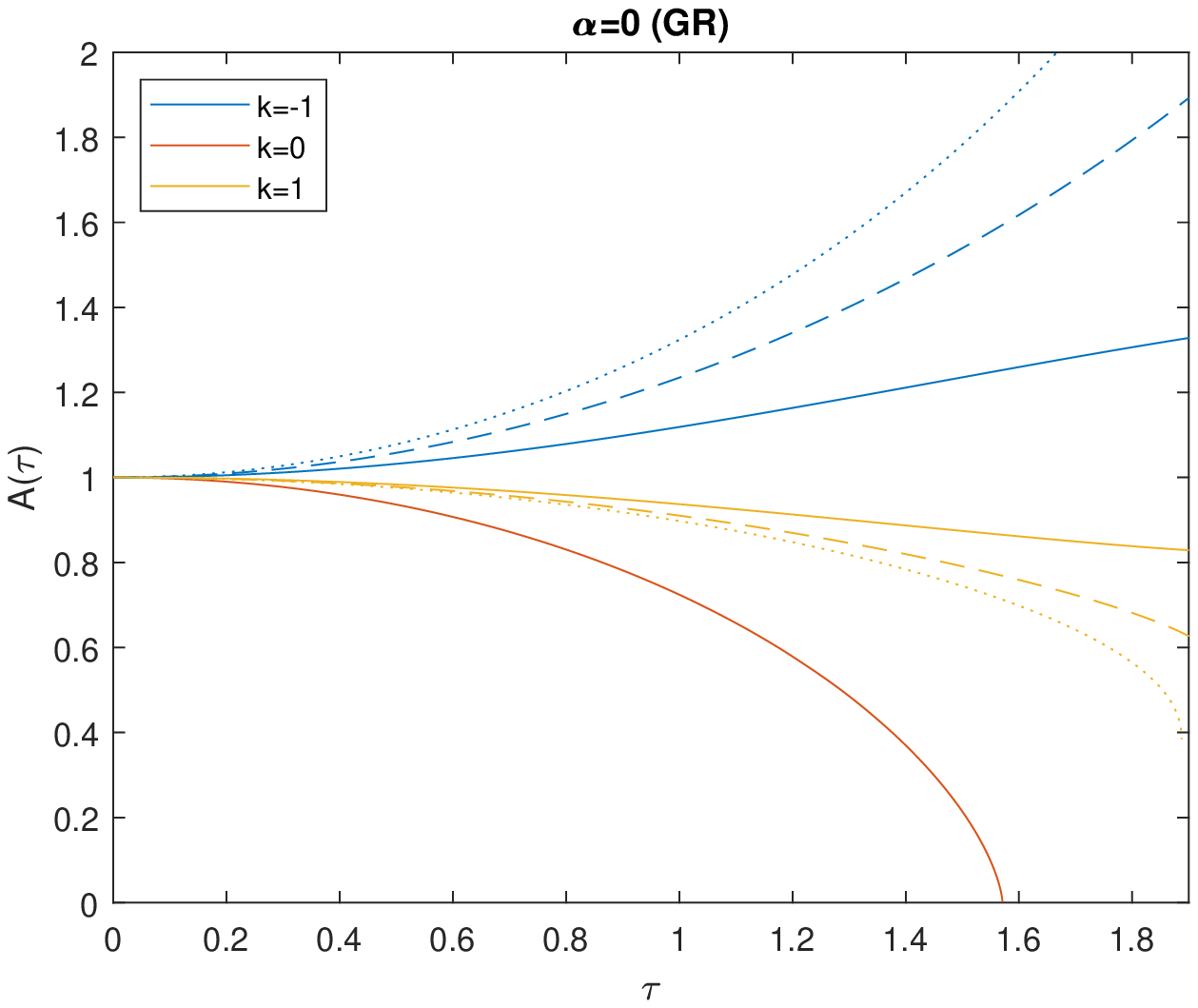}
  \caption{}
  \label{fig2b}
\end{subfigure}
\begin{subfigure}{.5\textwidth}\label{fig2c}
  \centering
 \includegraphics[scale=0.57]{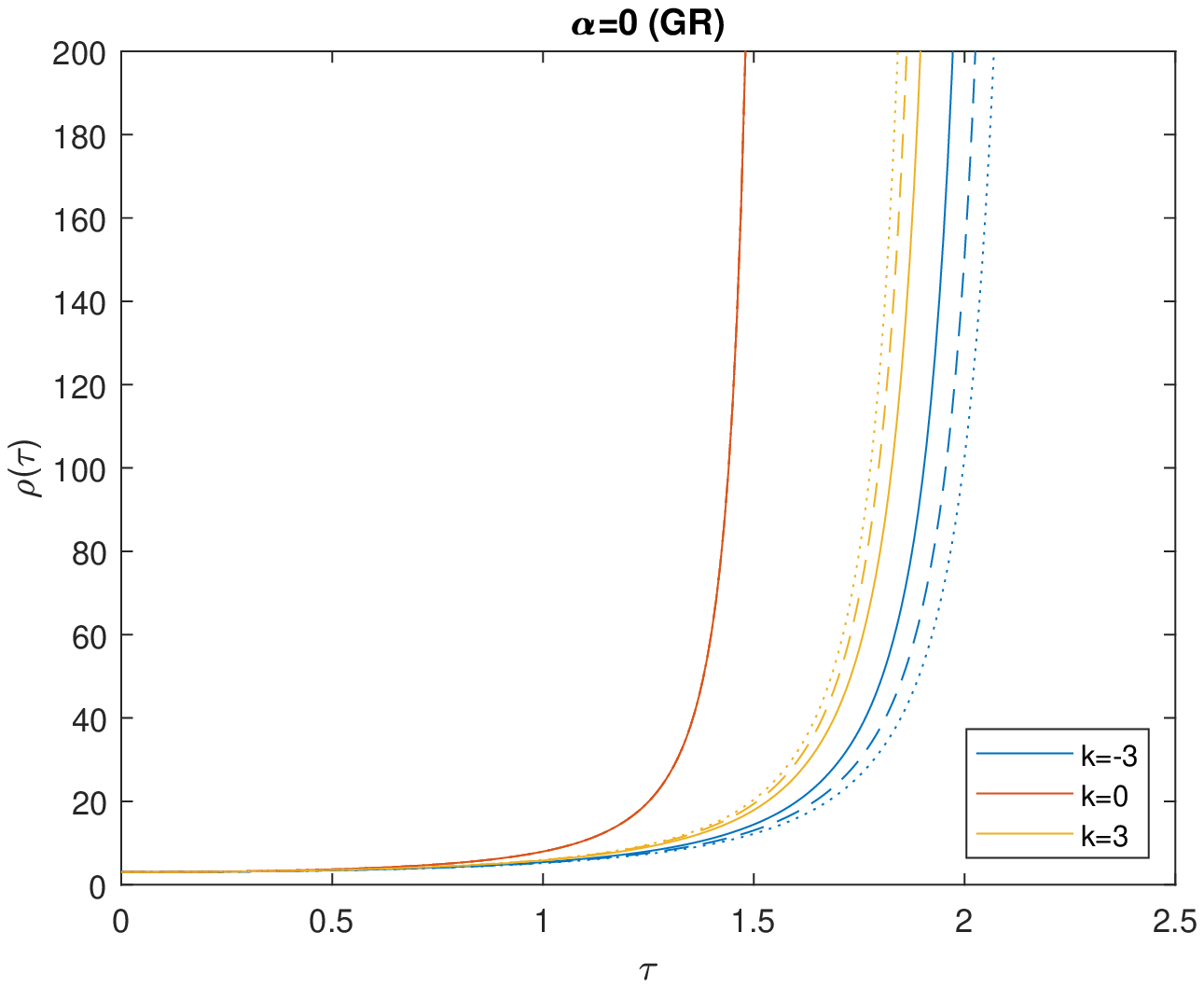}
  \caption{}
  \label{fig2c}
\end{subfigure}
\begin{subfigure}{.5\textwidth}\label{fig2d}
  \centering
 \includegraphics[scale=0.57]{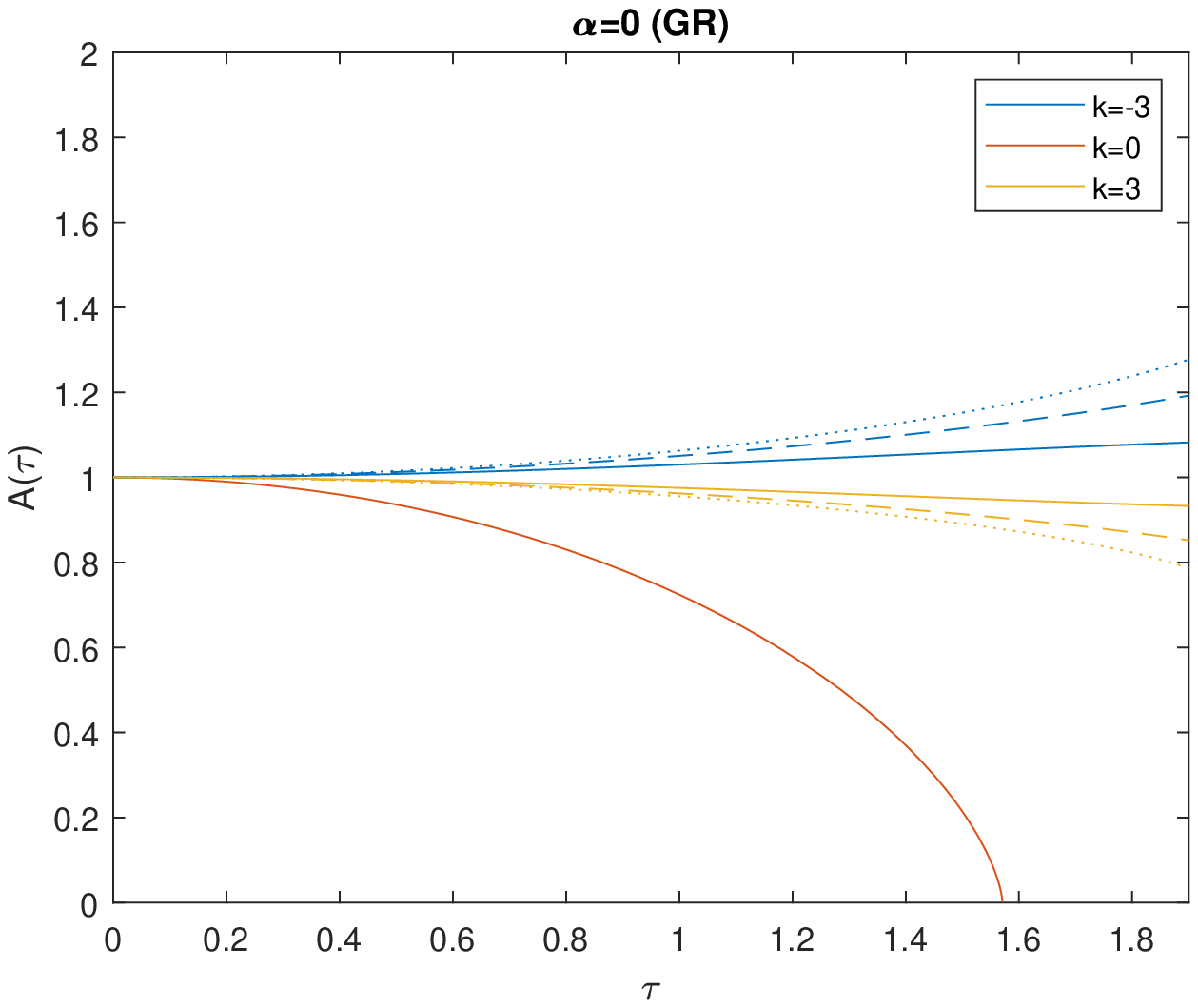}
  \caption{}
  \label{fig2d}
\end{subfigure}
\begin{subfigure}{.5\textwidth}\label{fig2e}
  \centering
 \includegraphics[scale=0.57]{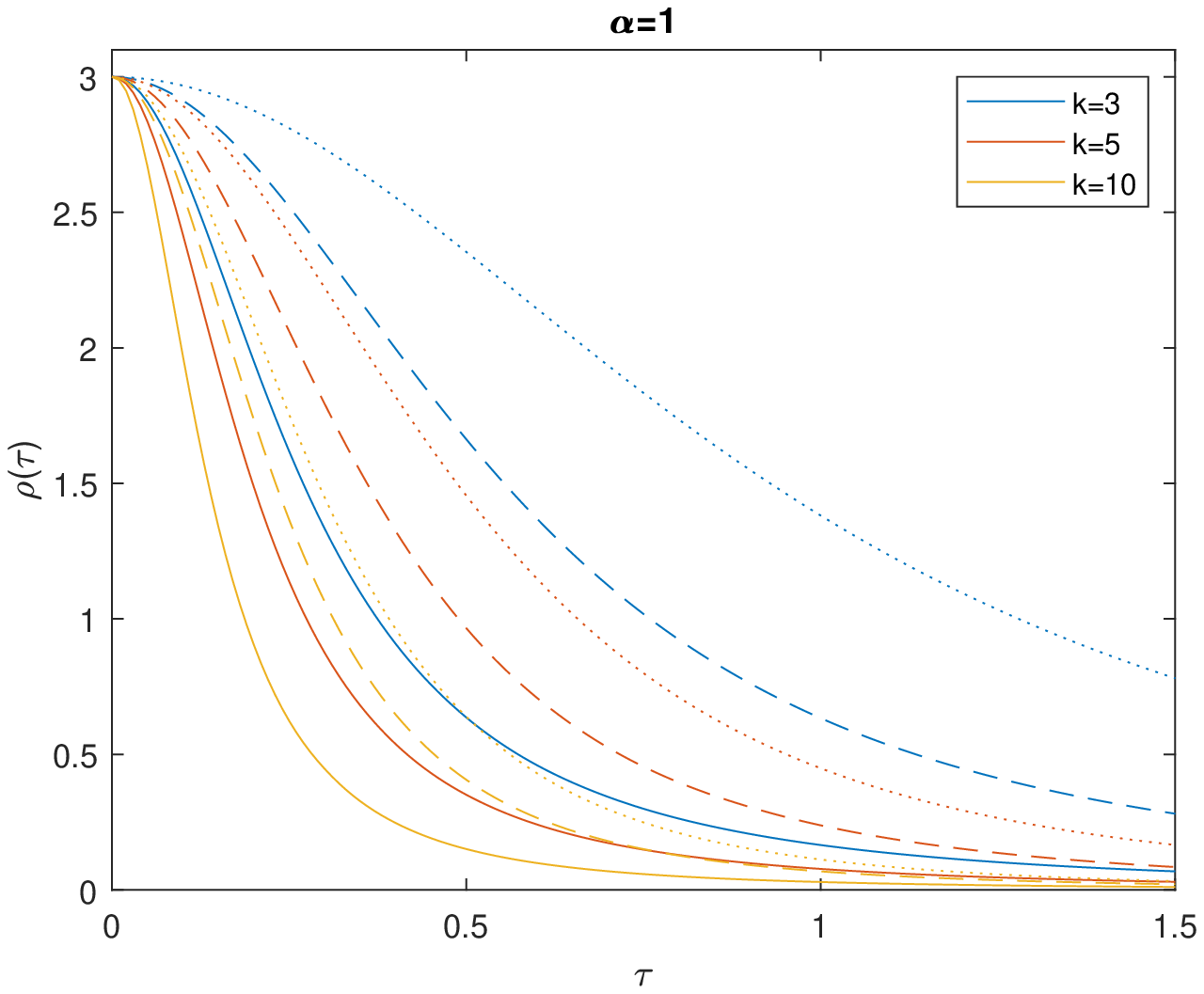}
  \caption{}
  \label{fig2e}
\end{subfigure}
\begin{subfigure}{.5\textwidth}\label{fig2f}
  \centering
 \includegraphics[scale=0.57]{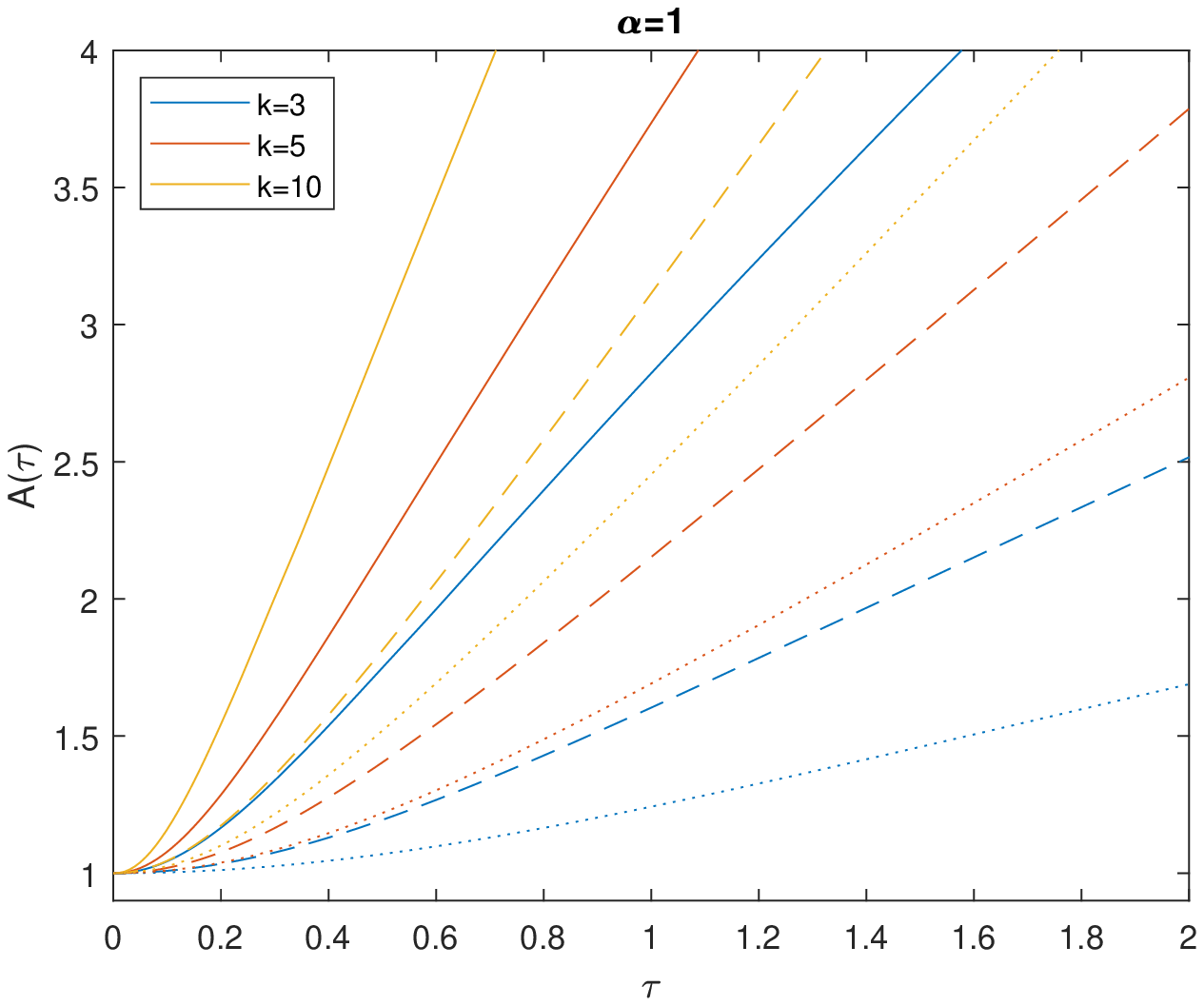}
  \caption{}
  \label{fig2f}
\end{subfigure}
\caption{Behavior of density (left panel) and scale factor (right panel) with increasing dimensionless time ($\tau$) for a polytrope  in the domain of modified gravity having Lagrangian $R+ 2\alpha T$. Here $n \in \{1.5,3,5\}$ and is represented by solid, dashed and dotted lines respectively. }\label{sf2}
\end{figure}

\begin{figure}
\ContinuedFloat
\begin{subfigure}{.5\textwidth}
  \centering
 \includegraphics[scale=0.57]{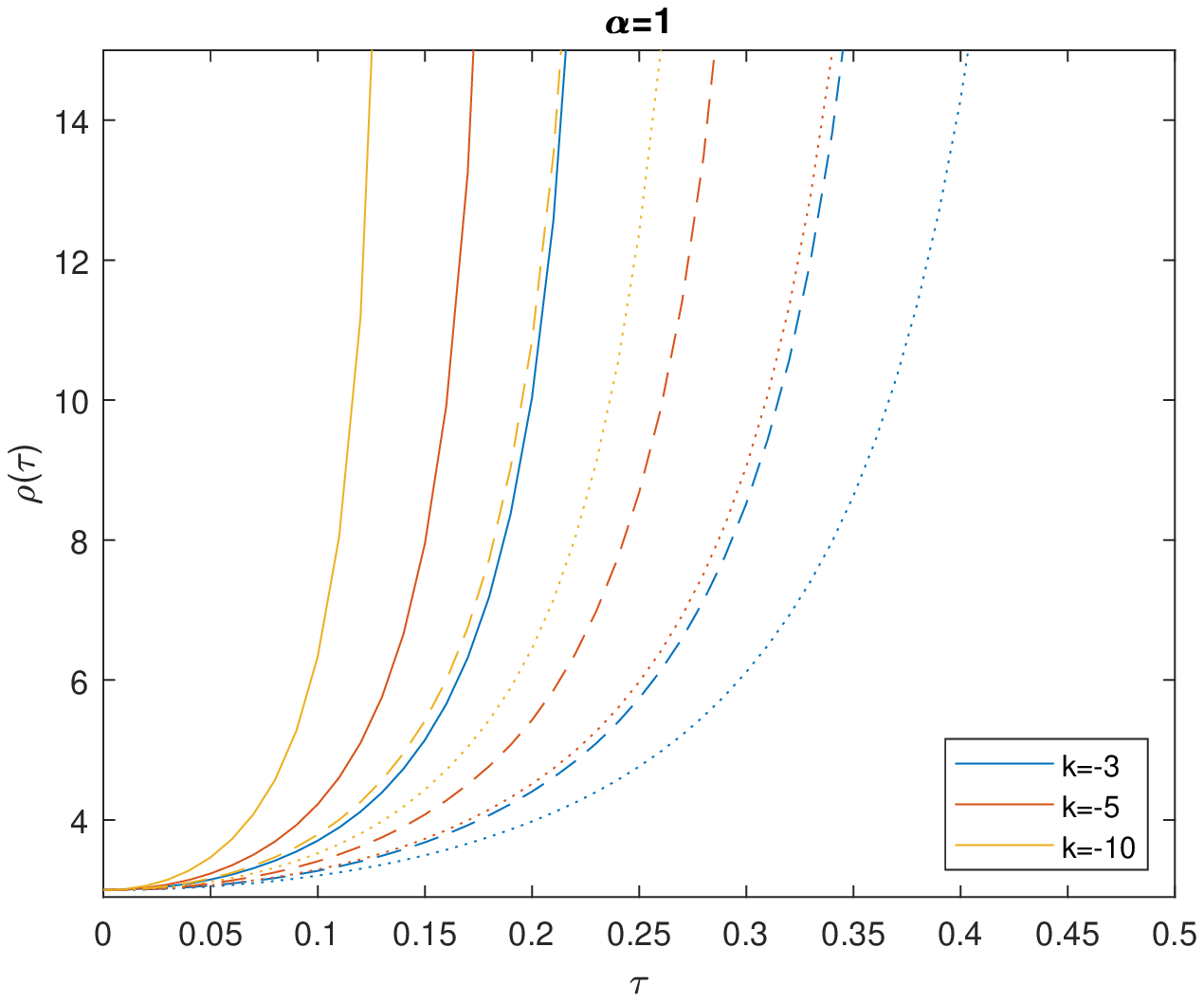}
  \caption{}
  \label{fig2g}
\end{subfigure}%
\begin{subfigure}{.5\textwidth}
  \centering
 \includegraphics[scale=0.57]{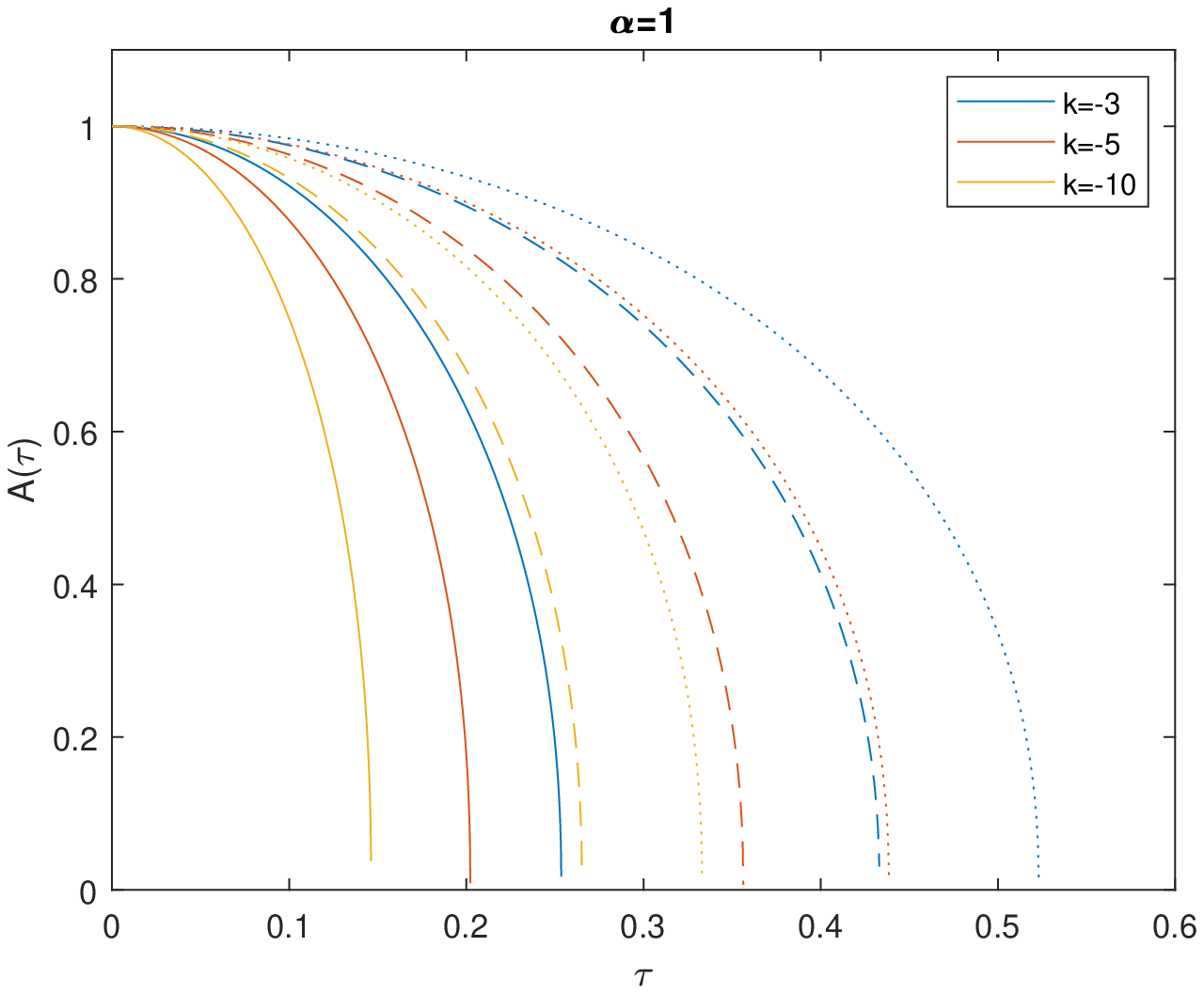}
  \caption{}
  \label{fig2h}
\end{subfigure}
\begin{subfigure}{.5\textwidth}
  \centering
 \includegraphics[scale=0.57]{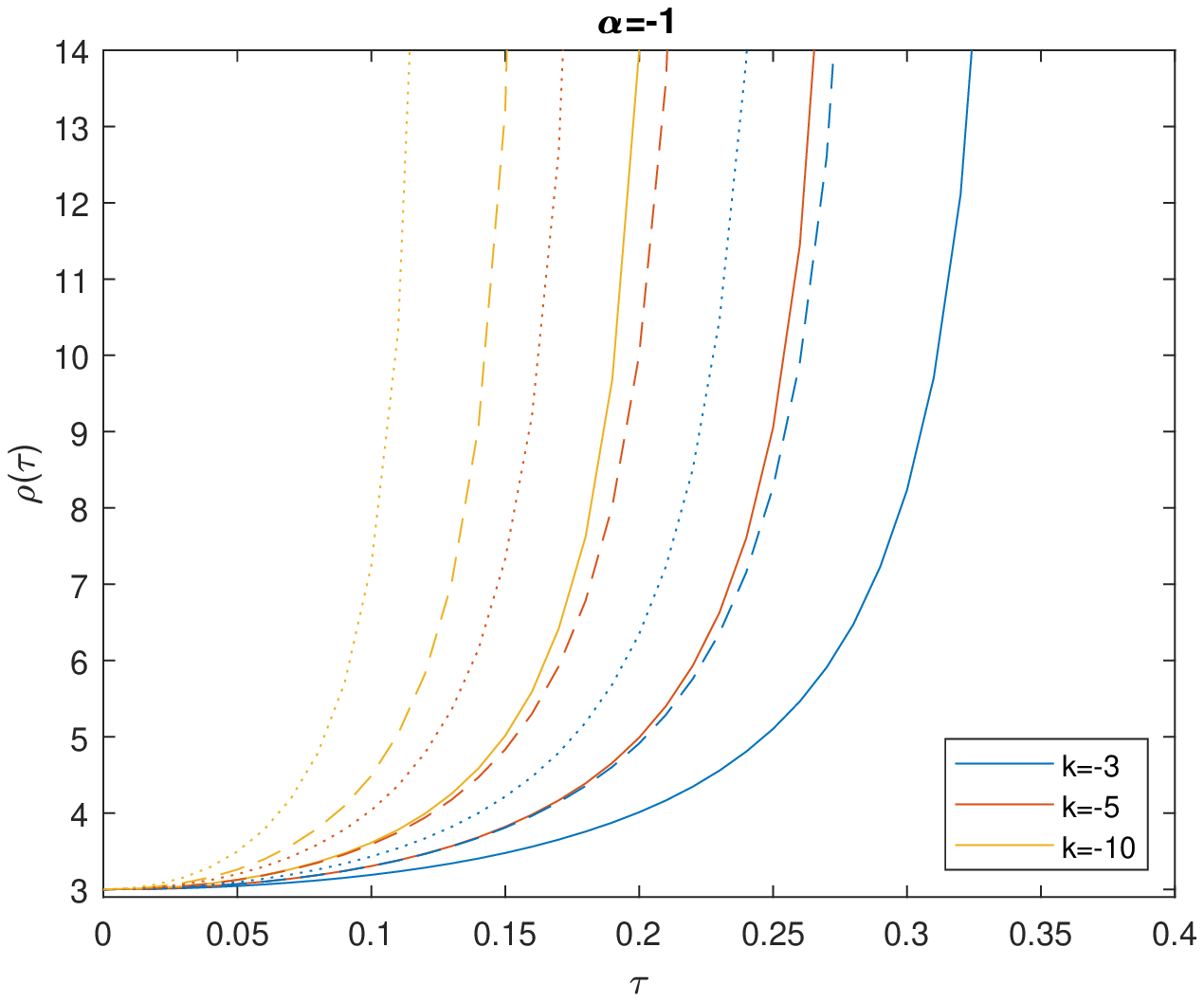}
  \caption{}
  \label{fig2i}
\end{subfigure}
\begin{subfigure}{.5\textwidth}
  \centering
 \includegraphics[scale=0.57]{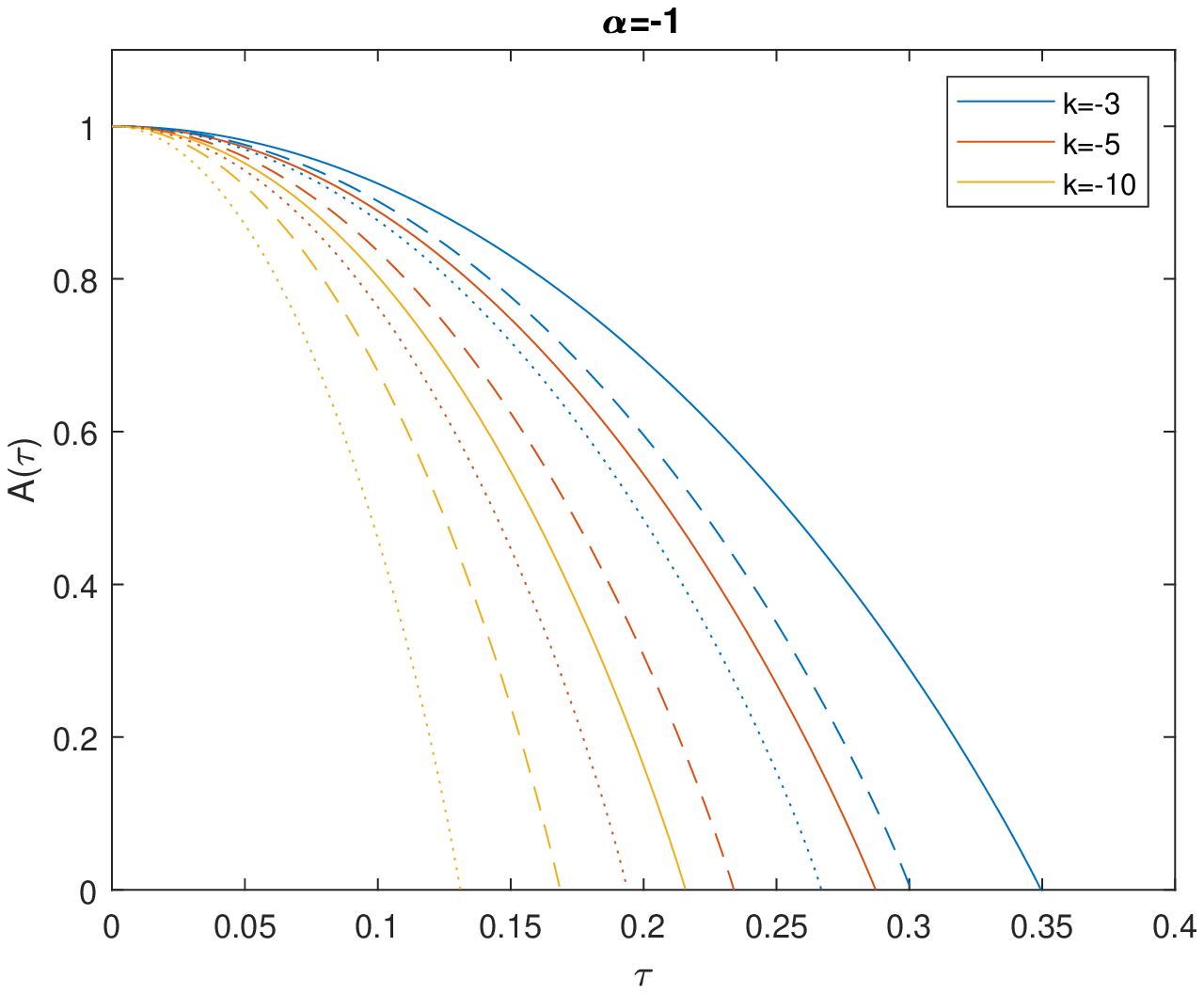}
  \caption{}
  \label{fig2j}
\end{subfigure}
\begin{subfigure}{.5\textwidth}
  \centering
 \includegraphics[scale=0.57]{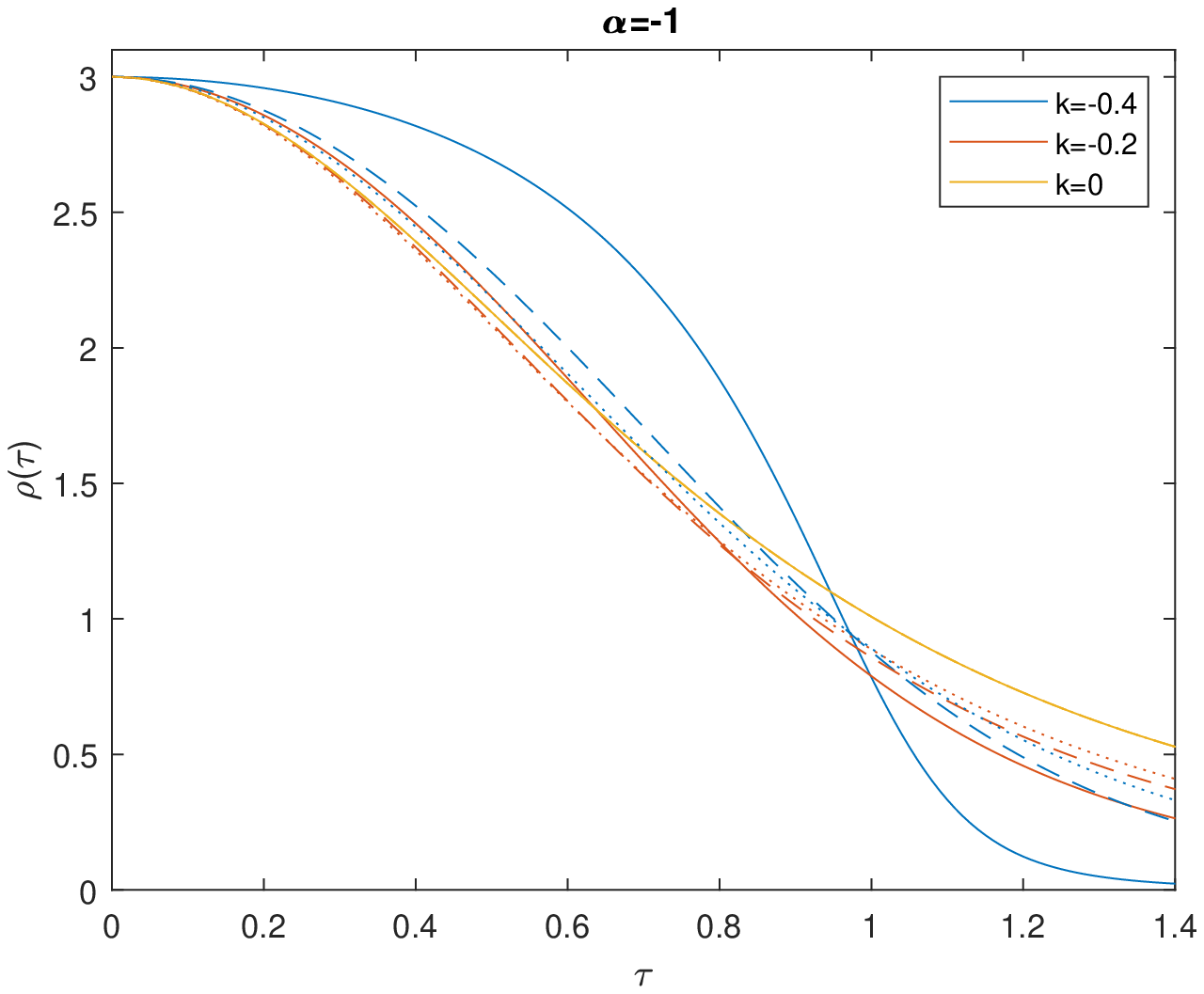}
  \caption{}
  \label{fig2k}
\end{subfigure}
\begin{subfigure}{.5\textwidth}
  \centering
 \includegraphics[scale=0.57]{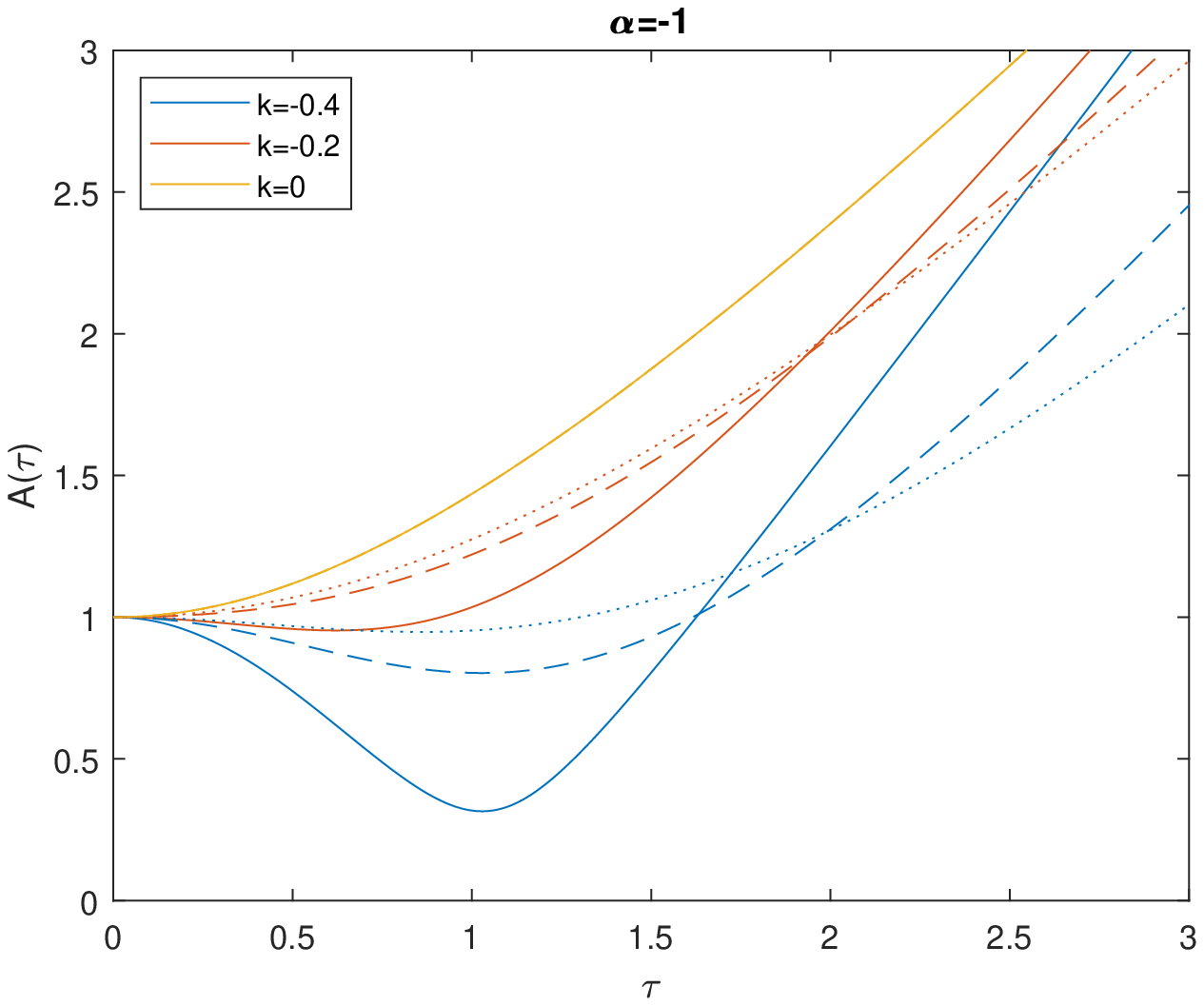}
  \caption{}
  \label{fig2l}
\end{subfigure}
\caption{Behavior of density (left panel) and scale factor (right panel) with increasing dimensionless time ($\tau$) for a polytrope  in the domain of modified gravity having Lagrangian $R+ 2\alpha T$. Here $n \in \{1.5,3,5\}$ and is represented by solid, dashed and dotted lines respectively. }\label{sf2}

\end{figure}

\begin{figure}
\ContinuedFloat
\begin{subfigure}{.5\textwidth}
  \centering
 \includegraphics[scale=0.57]{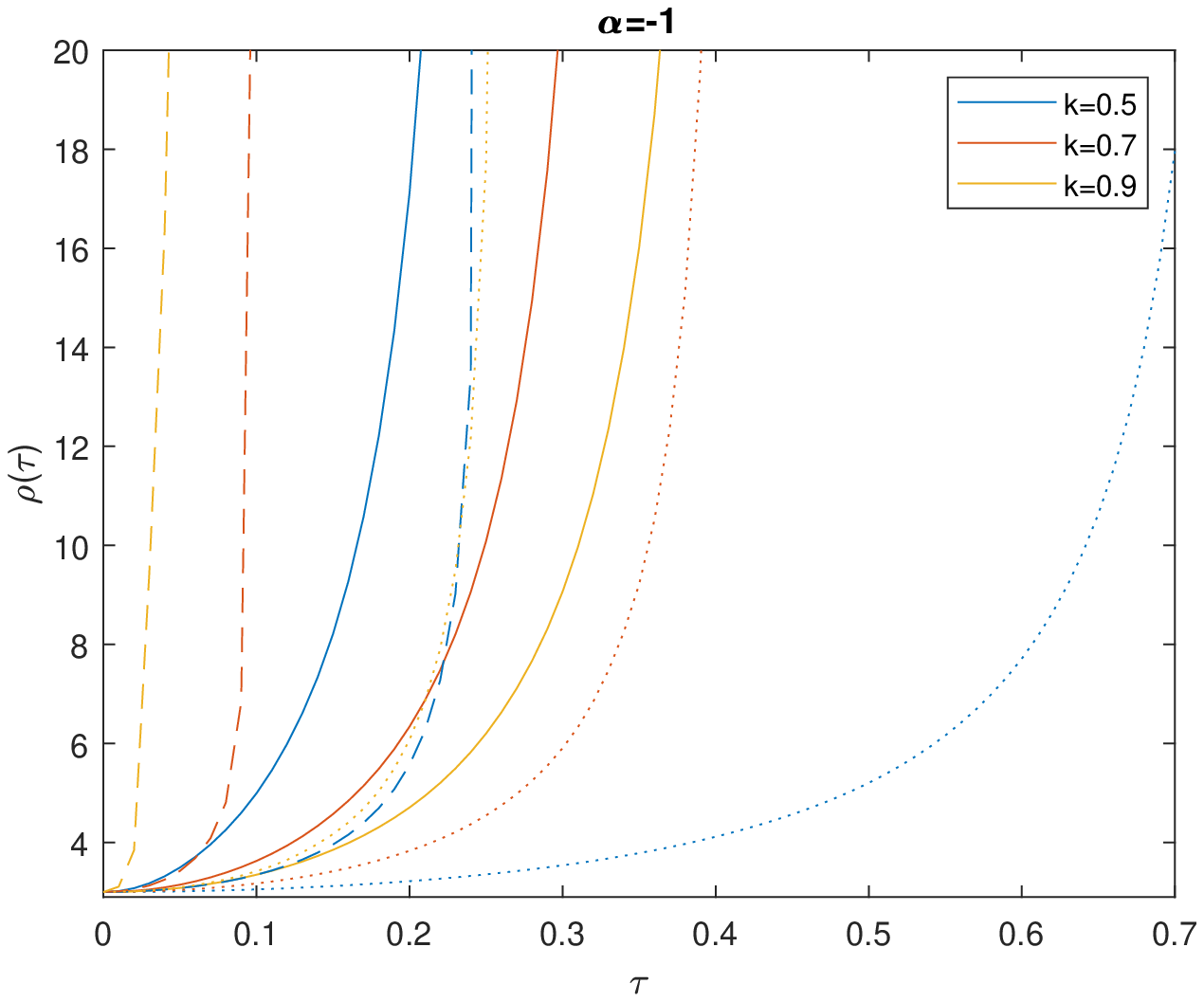}
  \caption{}
  \label{fig2m}
\end{subfigure}%
\begin{subfigure}{.5\textwidth}
  \centering
 \includegraphics[scale=0.57]{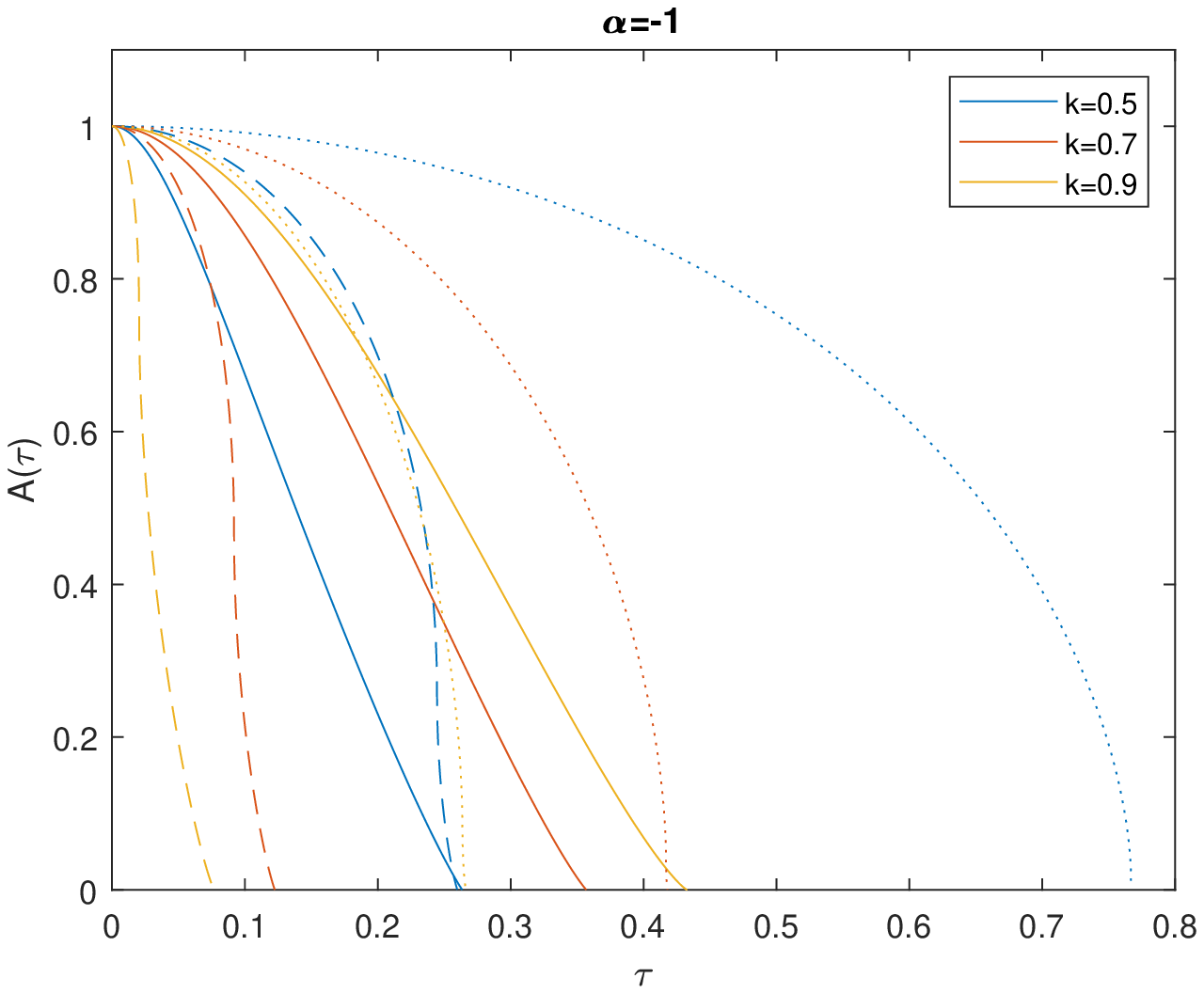}
  \caption{}
  \label{fig2n}
\end{subfigure}
\caption{Behavior of density (left panel) and scale factor (right panel) with increasing dimensionless time ($\tau$) for a polytrope  in the domain of modified gravity having Lagrangian $R+ 2\alpha T$. Here $n \in \{1.5,3,5\}$ and is represented by solid, dashed and dotted lines respectively. }\label{sf2}
\end{figure}
\begin{figure}
    \centering
    \includegraphics[scale=0.57]{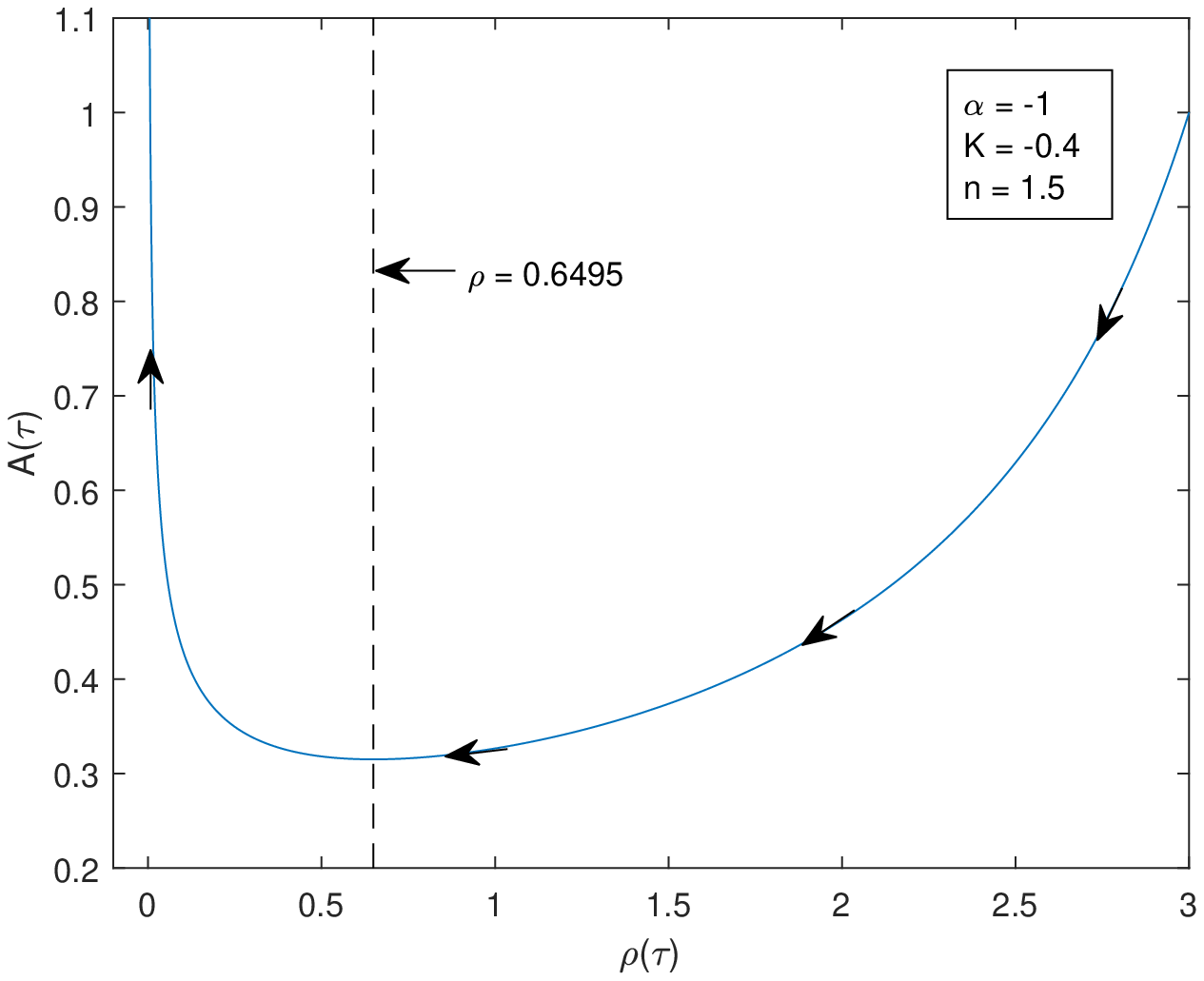}
    \caption{Parametric plot of scale factor ($A$) and density ($\rho$) for a particular set of variables, for a polytrope $p=K\rho^{1+\frac{1}{n}}$, in the domain of modified gravity having Lagrangian $R+2\alpha T$, with parameter being the dimensionless time ($\tau$) increasing along the arrow. }
    \label{fig3} 
\end{figure}
\subsection{Observations}
Since we have two differential equations governing the dynamics of density, differing in polarity, we choose the one which gives a real solution and not a complex one with non zero imaginary component. Dynamics of the density  and the corresponding dynamics of the scale factor has been plotted in Figure \ref{sf2}. Polytropic index of $1.5$ can approximate red giants, brown dwarfs and light white dwarfs while ploytropic index $3$ represents heavy white drwarfs and main sequence stars \cite{polytropicindex, polytropicindex2}.  The analysis is as follows: 
\begin{itemize}
 \item \textbf{$\alpha=0$ :} 
 
 The model of gravity corresponding to $\alpha=0$ is GR. Real values of $\rho$ are obtained when $\dot \rho>0$. Density becomes infinite in a finite time $\tau$. Parallel behavior of scale factor depends on the magnitude of $n$ and magnitude and polarity of $K$.
 
 For $K<0$, taking the limit of Eq.(\ref{eoc2}) as $\rho \to \infty$, which is obtained in a finite time in future, we get $A \to A_0$ ($<\infty$)  just like a ``Big Freeze" or type III singularity \cite{Bouhmadi} . In this form of singularity, the fluid system achieves a state of maximum entropy and the temperature reaches the absolute zero. Time of formation of singularity is more for larger value of polytropic index. It is worth noting that none of the energy conditions are satisfied. Since the density blows up in finite time, this behavior is analogous to a stellar model being infinitely filled with phantom energy ($\omega<-1$ corresponding to linear EoS).
 
 Similarly for positive value of $K$, $A \to 0$ as $\rho \to \infty$ in a finite time and a singularity analogous to  ``Big Crunch" or type 0 singularity \cite{Jambrina} is obtained. Increasing the polytropic index fastens the time of formation of singularity. Also increasing the magnitude of $K$ increases $\dot A$ and reduces the time of singularity formation. Apart from the Dominant Energy Condition (DEC), all the remaining energy conditions hold. If compared to the barotropic linear EoS, we can make this fluid behavior analogous to ordinary matter with linear EoS parameter $\omega >1$, thereby violating the causalty condition as well \cite{stiff}.
 \item \textbf{$\alpha=1$ :}
 
 %Numerically it is obtained that there is no real solution for any of the two differential Eq. (\ref{polytropecollapsedynamics}) for the values of $K_{1.5} \in (0,1.673)$, $K_3 \in (0,2.5305)$ and $K_5 \in (0.00879,2.7432)$, where the sub-script of $K$ is the corresponding polytropic index $n$. For values of $K$ above the upper limit and below the lower limit, real solutions are obtained when $\dot \rho <0$ and $\dot \rho >0$ respectively.

  From Figure \ref{fig2e} and Figure \ref{fig2f}, analysis for $K \geq 3$ has been done.  $\dot \rho<0$ for getting real solution for the Eq.(\ref{polytropecollapsedynamics}). The density vanishes at an infinite time and the fluid expands forever. Increasing $n$ slows down the expansion rate whereas rising $K$ fastens it. Coming to the discussion about the energy conditions, all but the DEC are satisfied. For positive $K$, the condition for DEC to hold true is that $K \leq 3^{-\frac{1}{n}}$. However, this is out of the feasible region giving real solutions for $\rho$. This type of fluid does not have any known counterpart or analog in the linear EoS type of fluid in the framework of GR. It behaves like the quintessence ($-1<\omega<-\frac{1}{3}$) in GR except that the DEC is satisfied by quintessence but not by this model fluid. 
 
 From Figure \ref{fig2g} and Figure \ref{fig2h}, for $K \leq -3$,  $\dot \rho>0$, a singularity analogous to  "Big Crunch" \cite{Jambrina} is obtained as the scale factor vanishes with a parallel blow up of the density. The time of formation of singularity is delayed by increasing $n$, whereas it hastens with increase in magnitude of $K$. The fluid has a similarity in the behavior with that of ordinary matter with linear EoS in GR except that unlike ordinary matter, none of the energy conditions are satisfied.
 %Again, this fluid does not have a counterpart in perfect fluid following linear EoS in GR since the only linear fluid in GR satisfying none of the energy conditions is the phantom one and obviously its behavior in not similar to the fluid under consideration.
 
 \item \textbf{$\alpha=-1$ :}
 
 From Figure \ref{fig2i} and Figure \ref{fig2j}, for $K \leq -3$, $\dot \rho >0$, again we get a singularity like the previous case. The time of formation of singularity hastens with  increase in $n$ as well as magnitude of $K$. Validity of energy condition and the behavior of the fluid is same as above.
 
  From Figure \ref{fig2k} and Figure \ref{fig2l}, one could see that for some particular value of $K$, for e.g. $K=-0.2$ or say $K=-0.4$, $\rho \to 0$ as time increases, and for late time, the scale factor increases. However, in early time an anomalous behavior is observed. Fixing $K=-0.4$, it is noted that $A$ decreases with reduction in density initially, but changes its behavior after some time and starts increasing. It seems that the fluid is changing its intrinsic property itself. Such fluid with time dependent property mimics the role of a dynamic scalar field in GR. Figure \ref{fig3} is the parametric plot of density v/s scale factor for $K=-0.4$ and $n=1.5$. The dynamics could be divided into two phases: ($1$) the collapsing phase and, ($2$) the expanding phase, divided by a fine line at $\rho \sim 0.6495$. Behavior of the fluid in phase ($1$) is analogous to the reverse behavior of linear EoS model fluid with $\omega<-1$ (phantom fluid), since in the latter case, $\rho$ and $A$ increases with time. Also, for Null Energy Condition (NEC)  to hold true, we need $\rho \geq \frac{1}{[-K]^{n}}$and for Strong Energy Condition (SEC), we need $\rho \geq \frac{1}{[-3K]^n}$, which could never be true in phase ($2$) since $\rho \to 0$ as $\tau \to \infty$.    
 
 Refering to Figure \ref{fig2m} and Figure \ref{fig2n}, which is analysed for $K \in \{0.5,0.7,0.9 \}$, we observe that density blows up in finite time and correspondingly the scale factor vanishes. There is no monotonic dependence on the polytropic index independent of $K$ and vice versa as indicated in the figure. Also amongst all the energy conditions, DEC is not satisfied.
  \end{itemize}
  
  %\textbf{Note:}  
  %De Sitter universe is a vacuum solution of the field equations in GR assuming the presence of a positive cosmological constant. Anti De Sitter universe is different in the sense that the cosmological constant is negative.
  %Having a positive $\alpha$ and $K$ as in Figure \ref{fig2e}/\ref{fig2f} has a negative cosmological term initially, and with increase in $\tau$, $\Lambda$ increases and be
\section{Summary}
By considering a spherically symmetric metric with components having separable terms, we have derived the field equations governing the stellar fluid corresponding to the Lagrangian in the EH action, $R+2\alpha T$, which is a candidate model to explain cosmic acceleration \cite{Harko1, Poplawski}. This model has equivalence with a GR model plus a cosmological term which is dynamic in nature, i.e. $\Lambda \propto H^2$. This is in agreement with the cosmological data favouring the idea of the cosmological term having variation with time. Studying the role of the constant term $\alpha \in \mathbb{R}$ reveals that in case of collapse of dust, radiation and stiff fluid, greater value of $\alpha$ hastens the collapse, and in case of expansion, more $\alpha$ slows down the expansion process. The turning point for all the three fluids from collapse to expansion are $\alpha =-\frac{1}{3}$, $-\frac{3}{4}$ and $-1$ respectively.

For general perfect fluid having linear EoS $p=\omega \rho$, the fluid having EoS parameter $\omega=\frac{3\alpha+1}{\alpha}$ or $\omega=\frac{2\alpha+1}{2\alpha+3}$ is static provided gravity is the only force of influence. A necessary and sufficient condition for a singularity to be physical and not merely a mathematical pathology is that $\zeta \neq 0$ and $\eta<2$, Eq.(\ref{zetaeta}), violating which results in vanishing of the Ricci or the Krettchmann scalar as $A \to 0$. Whether or not a singularity, if formed, is visible depends on whether an AH forms before the formation of singularity. It is observed that the sufficient condition for the  formation of AH prior to the formation of singularity is to have $\delta^2<\eta$. In the case when $\delta^2=\eta$, we get a momentary visible singularity which means that both the AH and the singularity forms simultaneously relative to the observer on the hypersurface that is the AH.   

For a stellar fluid having polytropic type EoS, after deriving the EoC, we solve it and get the scale factor  as a function of density, which is then used to get two differential equations in $\rho$, differing in polarity. The one which gives the real solution is the one which governs the dynamics of $\rho$. Once the density dependence on $\tau$ is obtained, it is then used to obtain the scale factor dependence on $\tau$.
%In the domain of GR, for $K<0$, singularity equivalent to Big Freeze  is obtained while for $K>0$, singularity similar  to Big Crunch  is obtained
Singularities similar to ``Big Freeze" and ``Big Crunch" has been obtained for certain  values of $\alpha$ and $K$. 

The case which is needed to be highlighted corresponds to the constant of proportionality $K$ and the polytropic index $n$ in some neighbourhood of $-0.4$ and $1.5$ respectively,  in the domain of gravity having Lagrangian $R-2T$. Here the scale factor reduces with reduction in the density of stellar fluid in its initial epoch. This behavior is analogous to time reversed behavior of the phantom fluid or it could simply be called ``Anti-Phantom Fluid". After achieving the minimum scale factor, the dynamics proceeds with increasing $A$ along with decreasing $\rho$. This  kaleidoscopic dynamics of $A$ is similar to the bouncing cosmological model \cite{Novello} where the singularity is avoided due to the presence of some exotic matter causing the universe to expand in later epochs after achieving the minimum scale factor. This model of polytropic fluid in presence of $\Lambda=-2T$ cosmological term if acted as a dominating fluid in the universe  gives support to the bouncing cosmology model.

%It is worth noting that the dynamics of both the linear as well as polytropic barotropic fluid has been investigated without invoking any matching conditions. 

\section{Acknowledgement}

The author Karim Mosani wishes to acknowledge the financial support (Junior Research Fellowship) provided by Council of Scientific and Industrial Research  (CSIR, India) File No.:09/919(0031)/2017-EMR-1 for carrying out this research work.
{}

 \end{document}